\documentstyle[epsf]{elsart}
\begin{document}

\begin{frontmatter}

\title{Realistic Models of Biological Motion}
\author{Imre Der\'enyi}
and
\author{Tam\'as Vicsek}
\address{Department of Atomic Physics, E\"otv\"os University\\
  Budapest, Puskin u 5-7, 1088 Hungary}

\begin{abstract}
The origin of biological motion can be traced back to the function of
{\em molecular motor proteins}. Cytoplasmic dynein and kinesin transport
organelles within our cells moving along a polymeric filament, the
microtubule.  The motion of the myosin molecules along the actin
filaments is responsible for the contraction of our muscles.  Recent
experiments have been able to reveal some important features of the
motion of individual motor proteins, and a new statistical physical
description -- often referred to as ``thermal ratchets'' -- has been
developed for the description of motion of these molecules.  In this
approach the motors are considered as Brownian particles moving along
one-dimensional periodic structures due to the effect of nonequilibrium
fluctuations.  Assuming specific types of interaction between the
particles the models can be made more realistic.  We have been able to
give analytic solutions for our model of kinesin with elastically
coupled Brownian heads and for the motion of the myosin filament where
the motors are connected through a rigid backbone.  Our {\em
theoretical predictions} are in a {\em very good agreement} with the
various {\em experimental results}.  In addition, we have considered
the effects arising as a result of interaction among a large number of
molecular motors, leading to a number of novel cooperative transport
phenomena.
\end{abstract}
\end{frontmatter}

\section{Introduction}
\label{s_intr}

The most common and best known transport phenomena occur in systems in
which there exist macroscopic driving forces (typically due to external
fields or concentration gradients).  However, recent theoretical
studies have shown that far from equilibrium processes in structures
possessing vectorial symmetry can bias thermal noise and induce
macroscopic motion on the basis of purely microscopic effects
\cite{aj_pr92,mag93,as_bi94,do_ho94,pr_ch94,aj_mu94,pe_er94,ba_ha94,mi_dy94,mil95,as_bi96,ha_ba96}.
This newly suggested mechanism is expected to be essential for the
operation of molecular combustion motors that are responsible for many
kinds of biological motion such as cellular transport or muscle
contraction \cite{lev95}.  In these cases motor proteins (dynein,
kinesin and myosin) convert the chemical energy stored in ATP molecules
into mechanical work while moving along polymeric filaments (dynein and
kinesin along microtubules, and myosin along actin filaments)
\cite{da_lo90}.  A transport mechanism of this kind has also been
experimentally demonstrated in simple physical systems
\cite{ro_sa94,fa_bo95,fa_li95} and can lead to new technological ideas
such as constructing nano-scale devices or novel type of particle
separators.

\begin{figure}[t]
\centerline{\epsfysize=5.2cm \epsfbox{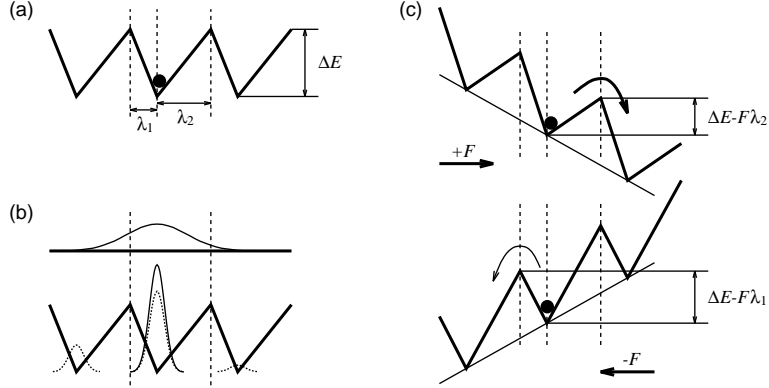}}
\caption{
The motion of a Brownian particle (a) in a sawtooth shaped potential
due to the effect of (b) fluctuating potential and (c) fluctuating
force.
}\label{fig1}\end{figure}

In the theoretical models -- loosely termed ``thermal ratchets'' --
molecular motors are considered as Brownian particles in an overdamped
environment moving along one-dimensional asymmetric periodic potentials
due to the effect of nonequilibrium fluctuations.  For simplicity and
illustration purposes let us choose the potential to be sawtooth shaped
(Fig.\ \ref{fig1}a), and consider the two basic types of fluctuations:
the fluctuating potential \cite{aj_pr92,as_bi94} (Fig.\
\ref{fig1}b) and the fluctuating force \cite{mag93,as_bi94} (Fig.\
\ref{fig1}c).

In the case of the fluctuating potential the sawtooth potential is
switched on and off repeatedly with appropriate switching rates.  Fig.\
\ref{fig1}b illustrates the time evolution of the probability
distribution of the position of the particle starting from one of the
potential valleys. If the temperature $T$ is small enough the particle
stays near the bottom of this valley until the potential is switched
off. Then the particle starts to diffuse on the flat potential.
Switching back to the sawtooth potential with the appropriate rate, the
particle has a large probability to fall back to the original valley or
to the neighboring valley to the left but only a small probability to
fall back to the right one because of the asymmetry of the potential.
Thus, a net motion to the left can be observed.

In the case of the fluctuating force $F(t)$
alternates between $+F$ and $-F$ with an appropriate characteristic
frequency and with zero time average (Fig.\ \ref{fig1}c).
When the force points to the
right ($F(t)=+F$), the probability of jumping to the right dominates
the jumping events with the coefficient $\exp(-(\Delta E -
F\lambda_2)/kT))$. When the force is $-F$, the probability of jumping
to the left dominates, but with the coefficient $\exp(-(\Delta E -
F\lambda_1)/kT))$. And since $\lambda_1<\lambda_2$, this fluctuating
force results in a net motion to the right.

These simple ratchet models catch the basic mechanism of the
operation of motor proteins, but they are too crude to be compared with
real experiments. However, assuming specific types of interaction
between the particles the models can be made more realistic.
We call a model of biological motion
realistic if it incorporates the relevant known features of motor
proteins and leads to predictions fully consistent with experiments.

\section{Kinesin walk model}
\label{s_kin}

Kinesin is a motor protein that moves large distances along microtubule
filaments and transports organelles and vesicles inside the cytoplasm
of eukaryotic cells. Native kinesin is a dimeric molecule with two
globular ($\sim$9$\times$3$\times3$~nm) heads. Microtubule is a long
hollow tube the wall of which is made up of tubulin heterodimers
arranged in 13 longitudinal rows called protofilaments. A tubulin
heterodimer is 8~nm long and consists of two globular proteins about
4~nm in diameter: $\alpha$- and $\beta$-tubulin. The dimers bind
head-to-tail giving the polarity to the protofilaments. Electron
micrograph measurements \cite{so_ma93,ho_sa95,ki_is95,hi_lo95} indicate
that kinesin heads can bind mainly to the $\beta$-tubulin.

Recent experimental studies in {\em in vitro} motility assays have
revealed numerous important properties of kinesin movement, such as the
unidirectionality of the motion, the force-velocity relationship, the
time dependence of the displacement and the displacement variance
\cite{va_re85,ra_me93,hu_gi94,sv_sc93,sv_bl94,sv_mi94}.

There are two basic classes of possible models for the stepping of
kinesin \cite{bl_sv95}.  The first one is the ``Long-Stride'' model in
which the heads are moving along a single protofilament.  The two heads
are displaced from each other by 8~nm.  During the stepping process the
back head passes the bound front head advancing 16~nm. Then the heads
change their roles and a new step may take place. This model was
studied in detail by Peskin and Oster \cite{pe_os95}. The other
possibility is the family of ``Two-Step'' models. These models
naturally explain the measured low displacement variance because of the two
sequential subprocesses: during one cycle one of the two heads takes a
8~nm step first then the other head steps. This kind of motion gives
large stability to the protofilament tracking. The differences in the
``Two-Step'' models arise from the relative positions of the heads.

\begin{figure}[t]
\centerline{\epsfysize=3.5cm \epsfbox{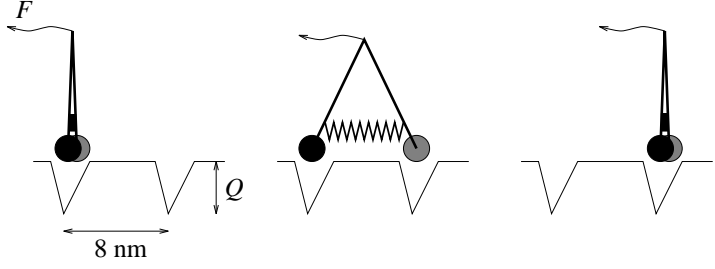}}
\caption{
Schematic picture of the potential and the subsequent steps of the
kinesin molecule.
}\label{fig2}\end{figure}

We propose a one-dimensional kinesin walk model \cite{de_vi96} which
describes the whole family of ``Two-Step'' models. Each of the two
Brownian heads of the kinesin can move along its own one-dimensional
periodic potential with period $L=8$~nm in an overdamped environment.
These potentials
represent the interaction with the protofilaments and the periods are
the tubulin heterodimers.  Each period has a deep potential valley
corresponding to the binding site of the $\beta$-tubulin and the other
parts of the potential are flat.  Each valley has an asymmetric ``V''
shape (see Fig.\ \ref{fig2}): the slope in the backward direction
is steeper and $0.5-1$~nm long, while the other slope is $1.5-2$~nm long.
These ranges are in the order of the Debye length.

The heads are connected at a hinge, and a {\em spring acts between
them}. The load force is joined to the heads through the hinge. At the
beginning of the mechano\-chemical cycle both heads are sitting in
their valleys waiting for an ATP mol\-ecule, and the spring is
unstrained. Any configuration of the model is mathematically equivalent
to a model in which the two potentials are identical, the heads are
sitting in the same valley, and therefore the rest length of the spring
is zero (Fig.\ \ref{fig2}).
After one of the heads binds an ATP molecule, the hydrolysis of this
ATP causes a conformational change in this head, more precisely,
induces the head to take an 8~nm forward step. In the language of the
model it means that the rest length of the spring changes from zero to
8~nm right after the hydrolysis. Then, as the first rate-limiting
subprocess, the strained spring is trying to stretch pushing the head
to the next valley. Reaching its new 8~nm rest length another
conformational change occurs in the head as a consequence of the ADP
release: the rest length of the spring changes back to zero quickly,
then, as the second rate-limiting subprocess, the spring is trying to
contract pulling forward the other head. Completing the contraction a
next cycle can start waiting for a new ATP molecule. This picture is
consistent with the scheme of Gilbert {\em et al.} \cite{gi_we95} for
the pathway of the kinesin ATPase, and with the very recent structural
data for the kinesin's head \cite{ku_sa96,sa_ku96}.
The parameters of the model (viscous drag coefficient, stiffness of the
spring, depth of the potential valleys) can be determined from well
known experimental data, therefore, they can only be tuned in a rather
restricted range.

\begin{figure}[t]
\centerline{\epsfxsize=4.8in \epsfbox{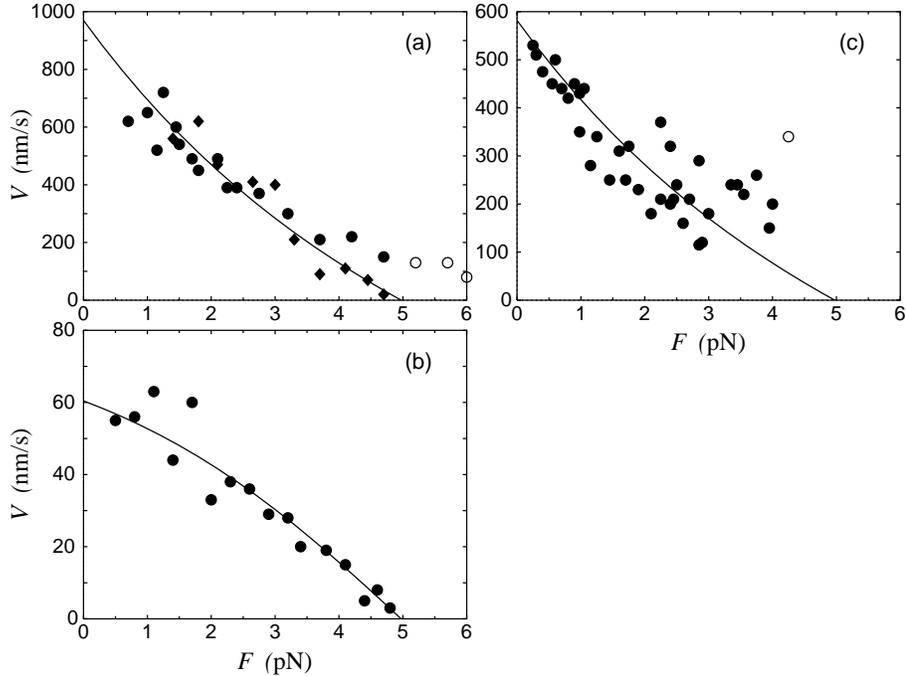}}
\caption{
The force-velocity curves for an individual kinesin molecule at
saturating ($\gg$ 90~$\mu$M) ATP in (a) and (c); and at low
($\sim$10~$\mu$M) ATP concentration in (b). The experimental data
(scattered symbols) are from Svoboda and Block \protect\cite{sv_bl94}
in (a) and (b); and from Hunt {\it et al.} \protect\cite{hu_gi94} in (c).
The open
circles correspond to the simultaneous effect of multiple kinesin
motors in the authors' interpretation \protect\cite{hu_gi94,sv_bl94}.
The circles and diamonds in (a) mean two different set of the measured
data.
}\label{fig3}\end{figure}

Due to the asymmetry of the potential under low load force $F$ the kinesin
molecule steps 8~nm forward during almost each mechanochemical cycle.
But increasing the load the probability that the molecule remains on
the same place or even takes a backward 8~nm step increases.
The analytic solution of the model \cite{de_vi96} results in a very
good fit to the measured mechanical properties of the kinesin walk
(see, e.g., the force-velocity relationship in Fig.\ \ref{fig3}).

\section{Muscle contraction}
\label{s_myo}

\begin{figure}[b]
\centerline{\epsfxsize=4.8in \epsfbox{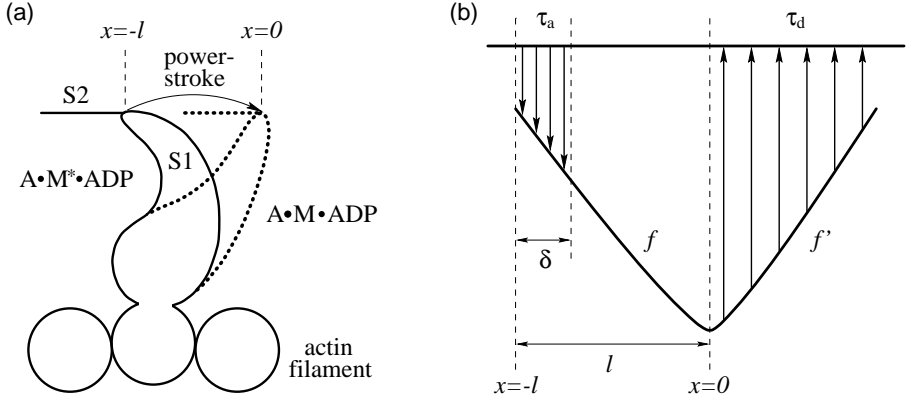}}
\caption{
Cartoon (a) shows the conformational change in myosin head during the
power-stroke. On the schematic picture (b) the flat and ``V'' shaped
potentials represent the weakly and strongly bound states of the myosin
cross-bridge. The slopes of the ``V'' shaped potential are $f$ and $f'$
that correspond to the force produced by the myosin head in the
power-stroke and in the drag-stroke region respectively. Transition
from the weakly bound state to the strongly bound state can occur in
the $-l<x<-l+\delta$ region with characteristic time $\tau_{\rm a}$ and
transition from the strongly bound state to the weakly bound state in
the $0<x$ with characteristic time $\tau_{\rm d}$. These time constants
are determined by the kinetic rates and the substrate concentrations.
}\label{fig4}\end{figure}

Muscle contraction results from a cyclic interaction between the actin
filaments and their motor proteins -- the myosin molecules. During the
contraction the myosin cross-bridges extending from the thick myosin
filaments attach to the binding sites of the thin actin filaments and
exert force on them yielding a relative sliding of the actin and myosin
filaments. The widely accepted scheme for the cross-bridge cycle
\cite{gol87,bag93}, which was originally proposed by Lymn and Taylor
\cite{ly_ta71}, is \\
\centerline{\epsfbox{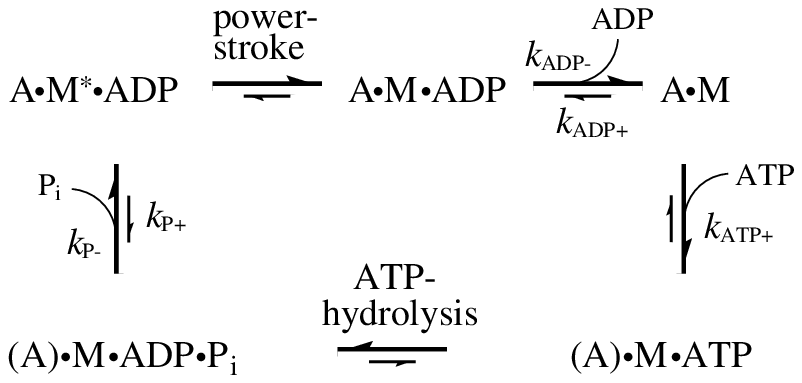}} \\
Initially, the myosin cross-bridge is in a weakly bound,
non-force-producing $\rm (A)\cdot M\cdot ADP\cdot P_i$ state. A and M
denote actin and myosin respectively, and (A) indicates that myosin is
weakly bound to actin. Releasing the phosphate $\rm P_i$ the cross-bridge
gets into a strongly bound, force-producing $\rm A\cdot M^*\cdot ADP$
state. Then the cross-bridge executes a {\em power stroke} (see Fig.\
\ref{fig4}a) producing mechanical force on actin, a conformational
change in myosin, and a relative sliding between adjacent actin and myosin
filaments. This new $\rm A\cdot M\cdot ADP$ state cannot readily bind
$\rm P_i$ any more, in contrast to the preceding one. At the end of the
power-stroke, when continued attachment would resist useful work, ADP is
released with transition to the $\rm A\cdot M$ state, which is still
strongly bound. ADP may then rebind, resulting in continued
counter-productive attachment, or ATP may bind, resulting in rapid
cross-bridge detachment. The hydrolysis step occurs in this weakly bound
$\rm (A)\cdot M\cdot ATP$ state resulting the $\rm (A)\cdot M\cdot
ADP\cdot P_i$ cross-bridge state, which is free to participate in a new
working cycle. The conformations of the
cross-bridge states was determined by EPR spectroscopy \cite{os_th95}.

One of the most important problems is to derive the macroscopic
mechanical properties of muscle contraction from the microscopic
biochemical steps of the cross-bridge interaction. The first and very
influential analysis was presented by Huxley \cite{hux57}. With his
original two-state model
he was able to reproduce much of the
experimental data available at that time. Since then the models have
become finer and more complicated to fit the increasing number of
experimental data
\cite{hil74,ei_hi80,pa_co89,pa_wh93,co_er92,le_hu93}.
The general disadvantages of these models are that
(i) they incorporate several functions among the other fitting parameters,
(ii) and they can be solved only numerically.

We propose a simple mechanochemical model \cite{de_vi97} for the
cross-bridge interaction that is free from the above mentioned
shortcomings. Our four-state model
(i) involves the {\em relevant} kinetic data and other microscopic
parameters,
(ii) and provides simple {\em analytic} solutions for the mechanical
properties of muscle contraction.
These analytic results fit numerous experimental data extremely well
and make clear the relationship between the microscopic and macroscopic
parameters.
We would like to point out that analytic solutions are extremely useful
if one would like to go beyond merely giving a phenomenological
description of a biological phenomenon, since the structure of the
analytic expressions explicitly relating the relevant factors provides
a unique insight into the role of these factors.
The model is based on the following assumptions:

(i) The distance between two neighboring binding sites on the actin
filament is $d\approx 37$~nm being the half period of the double helix
of the actin filament. The position of the actin filament is fixed
while the myosin filament passes by it with a velocity $V$ from left to
right. Since the spacing of the myosin heads and the actin binding
sites are incommensurate, the distances between the myosin
cross-bridges and their nearest binding sites are uniformly distributed
within the range $[-d/2,d/2]$. Thus, the macroscopic properties of
muscle contraction can be calculated as a spatial average for the
position of one cross-bridge (more precisely its S1-S2 junction,
denoted by $x$) over this interval.

(ii) Initially the myosin cross-bridge is in a weakly bound,
non-force-producing state that can
be represented by a flat potential (Fig.\ \ref{fig4}b).
When the cross-bridge gets into the vicinity $-l\leq x\leq
-l+\delta$ of the actin binding site, phosphate can be released
forming a stereospecific, strongly bound,
force-producing state.
$l$ denotes the power-stroke working distance for which the most
accurate value $l\approx 11$~nm was measured by using optical
tweezers \cite{fi_si94}. This is also consistent with the structural studies
of myosin \cite{ra_ho93,fi_sm95}.
$\delta$ is the composition of the Debye length and the rotational
freedom of the myosin head, and can be well approximated by a few nm.
The force produced by
this state in the power-stroke region ($-l\leq x\leq 0$) is about
$f\approx 4$~pN which was also measured by Finer {\it et al.}
\cite{fi_si94}. Finishing the power-stroke continued attachment resists
useful work, i.e., exhibits a negative force $-f'$ in this {\it
drag-stroke} region ($0<x$). $f'$ must be close to $f$, but not
necessarily equals to it. Altogether, the strongly bound state
can be represented by a ``V'' shaped potential shown in Fig.\
\ref{fig4}b.

(iii) In the drag-stroke region ($0<x$)
ADP can be released with transition to another strongly
bound state which is supposed to have similar properties, and can also
be represented by the same ``V'' shaped potential.
Then ATP can bind resulting in rapid cross-bridge detachment.
And finally a hydrolysis step occurs in this weakly bound state
which is represented by the flat potential.

\begin{figure}[t]
\centerline{\epsfxsize=4.8in \epsfbox{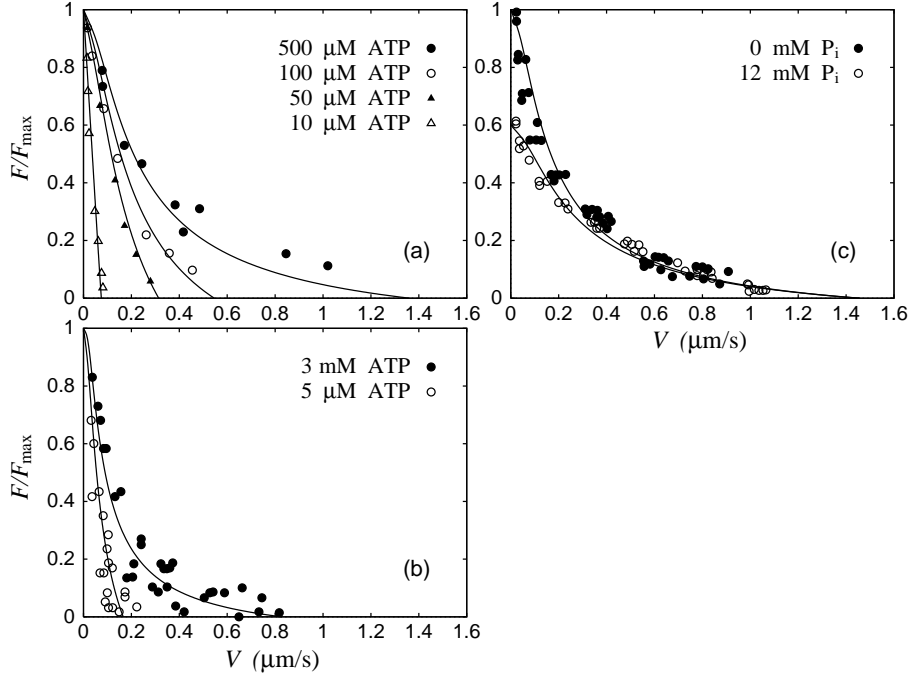}}
\caption{
The force-velocity relationship of the contracting muscle in various
situations: rabbit psoas muscle \protect\cite{co_bi79} in (a) and
rabbit semimembranosus muscle \protect\cite{pa_li92} in (b) for
different ATP concentrations in the absence of ADP and $\rm P_i$; and
rabbit psoas muscle \protect\cite{co_pa85} in (c) for two different
$P_i$ concentrations in the absence of ADP at saturating (4~mM) ATP.
The scattered symbols are experimental data and the solid lines are
fits of the model.
}\label{fig5}\end{figure}

The analytic solution of this model \cite{de_vi97}
agrees with numerous experimental results very well:
Fig.\ \ref{fig5} shows some fits to the measured force-velocity
\cite{co_bi79,pa_li92,co_pa85}. The well known
empirical Hill equation \cite{hil38} also gives a good fit,
although, its relationship to the
biochemical level is unknown.
The model results in a Michaelis-Menten-like saturation
behavior for the maximum shortening velocity
\cite{co_bi79,pa_li92,co_pa85,kr_sp86,ha_no87} and the ATPase activity
\cite{sl_gl86} as a function of the ATP concentration, which is also in
good agreement with the experimental results. Further features of our
model are also consistent with the experiments
\cite{kat39,hu_si71,ed_hw77}. In addition, the model gives a prediction
for the proportion of the attached cross-bridges. This is very
difficult to measure and the recent experimental results are
contradictory.


\section{Cooperative motors}
\label{s_coop}


From the previous sections it has turned out that specific interaction
between the motors has significant effect on their motion.  Elastic
interaction between two \cite{ajd94} or more \cite{cs_fa97} particles
and a general model of rigidly coupled motors \cite{ju_pr95} has been
recently investigated that revealed numerous peculiar collective
phenomena.

Since in real biological \cite{as_sc90} and physical \cite{ro_sa94}
systems a large number of free motors can move along the same structure
we have investigated the collective motion of finite sized Brownian
particles in a one-dimensional sawtooth shaped potential for both the
fluctuating force \cite{de_vi95} and the fluctuating barrier
\cite{de_aj96}. The interaction between the particles is supposed to be
a simple hard core repulsion, thus the particles can be considered as
hard rods. This type of interaction means that during the motion the
particles are not allowed to overlap.

Analytical treatment of some special sets of parameters and computer
simulations have revealed that this simple system exhibits many
interesting collective transport properties, e.g.:

(i) Increasing the density of the particles (defined as the number of
particles $\times$ the size of the particles / the system size) from 0
to 1 the direction of their motion can change its sign several times.

(ii) Close to the maximal density the the average velocity shows a very
strong and complex dependence on the size of the particles.

\bibliography{paper}

\begin{thebibliography}{10}

\bibitem{aj_pr92}
A.~Ajdari and J.~Prost.
\newblock {\em C. R. Acad. Sci. Paris}, 315:1635, 1992.

\bibitem{mag93}
M.~O. Magnasco.
\newblock {\em Phys. Rev. Lett.}, 71:1477, 1993.

\bibitem{as_bi94}
R.~D. Astumian and M.~Bier.
\newblock {\em Phys. Rev. Lett.}, 72:1766, 1994.

\bibitem{do_ho94}
C.~R. Doering, W.~Horsthemke, and J.~Riordan.
\newblock {\em Phys. Rev. Lett.}, 72:2984, 1994.

\bibitem{pr_ch94}
J.~Prost, J.-F. Chauwin, L.~Peliti, and A.~Ajdari.
\newblock {\em Phys. Rev. Lett.}, 72:2652, 1994.

\bibitem{aj_mu94}
A.~Ajdari, D.~Mukamel, L.~Peliti, and J.~Prost.
\newblock {\em J. Phys. I France}, 4:1551, 1994.

\bibitem{pe_er94}
C.~S. Peskin, G.~B. Ermentrout, and G.~F. Oster.
\newblock In V.~C. Mow et~al., editors, {\em Cell Mechanics and Cellular
  Engineering}. Springer-Verlag, New York, 1994.

\bibitem{ba_ha94}
R.~Bartussek, P.~H{\"a}nggi, and J.~G. Kissner.
\newblock {\em Europhys. Lett.}, 28:459, 1994.

\bibitem{mi_dy94}
M.~M. Millonas and D.~I. Dykman.
\newblock {\em Phys. Lett. A}, 185:65, 1994.

\bibitem{mil95}
M.~M. Millonas.
\newblock {\em Phys. Rev. Lett.}, 74:10, 1995.

\bibitem{as_bi96}
R.~D. Astumian and M.~Bier.
\newblock {\em Biophys. J.}, 70:637, 1996.

\bibitem{ha_ba96}
P.~H{\"a}nggi and R.~Bartussek.
\newblock In J.~Parisi, S.~C. M{\"u}ller, and W.~Zimmermann, editors, {\em
  Nonlinear Physics of Complex Systems, Lecture Notes in Physics}, volume 467.
  Springer, Berlin, 1996.

\bibitem{lev95}
B.~G. Levi.
\newblock {\em Phys. Today}, 48(4):17, 1995.

\bibitem{da_lo90}
J.~Darnell, H.~Lodish, and D.~Baltimore.
\newblock {\em Molecular Cell Biology}.
\newblock Scientific American Books, New York, 1990.

\bibitem{ro_sa94}
J.~Rousselet, L.~Salome, A.~Ajdari, and J.~Prost.
\newblock {\em Nature}, 370:446, 1994.

\bibitem{fa_bo95}
L.~P. Faucheux, L.~S. Bourdieu, P.~D. Kaplan, and A.~J. Libchaber.
\newblock {\em Phys. Rev. Lett.}, 74:1504, 1995.

\bibitem{fa_li95}
L.~P. Faucheux and A.~J. Libchaber.
\newblock {\em J. Chem. Soc. Faraday. Trans.}, 91:1363, 1995.

\bibitem{so_ma93}
Y.-H. Song and E.~Mandelkow.
\newblock {\em Proc. Natl. Acad. Sci. USA}, 90:1671, 1993.

\bibitem{ho_sa95}
A.~Hoenger, E.~P. Sablin, R.~D. Vale, R.~J. Fletterick, and R.~A. Milligan.
\newblock {\em Nature}, 376:271, 1995.

\bibitem{ki_is95}
M.~Kikkawa, T.~Ishikawa, T.~Wakabayashi, and N.~Hirokawa.
\newblock {\em Nature}, 376:274, 1995.

\bibitem{hi_lo95}
K.~Hirose, A.~Lockhart, R.~A. Cross, and L.~A. Amos.
\newblock {\em Nature}, 376:277, 1995.

\bibitem{va_re85}
R.~D. Vale, T.~S. Reese, and M.~P. Sheetz.
\newblock {\em Cell}, 42:39, 1985.

\bibitem{ra_me93}
S.~Ray, E.~Meyh{\"o}fer, R.~A. Milligan, and J.~Howard.
\newblock {\em J. Cell. Biol.}, 121:1083, 1993.

\bibitem{hu_gi94}
A.~J. Hunt, F.~Gittes, and J.~Howard.
\newblock {\em Biophys J.}, 67:766, 1994.

\bibitem{sv_sc93}
K.~Svoboda, C.~F. Schmidt, B.~J. Schnapp, and S.~M. Block.
\newblock {\em Nature}, 365:721, 1993.

\bibitem{sv_bl94}
K.~Svoboda and S.~M. Block.
\newblock {\em Cell}, 77:773, 1994.

\bibitem{sv_mi94}
K.~Svoboda, P.~P. Mitra, and S.~M. Block.
\newblock {\em Proc. Natl. Acad. Sci. USA}, 91:11782, 1994.

\bibitem{bl_sv95}
S.~M. Block and K.~Svoboda.
\newblock {\em Biophys J.}, 68:230s, 1995.

\bibitem{pe_os95}
C.~S. Peskin and G.~Oster.
\newblock {\em Biophys J.}, 68:202s, 1995.

\bibitem{de_vi96}
I.~Der\'enyi and T.~Vicsek.
\newblock {\em Proc. Natl. Acad. Sci. USA}, 93:6775, 1996.

\bibitem{gi_we95}
S.~P. Gilbert, M.~R. Webb, M.~Brune, and K.~A. Johnson.
\newblock {\em Nature}, 373:671, 1995.

\bibitem{ku_sa96}
F.~J. Kull, E.~P. Sablin, R.~Lau, R.~J. Fletterick, and R.~D. Vale.
\newblock {\em Nature}, 380:550, 1996.

\bibitem{sa_ku96}
E.~P. Sablin, F.~J. Kull, R.~Cooke, R.~D. Vale, and R.~J. Fletterick.
\newblock {\em Nature}, 380:555, 1996.

\bibitem{gol87}
Y.~E. Goldman.
\newblock {\em Ann. Rev. Physiol.}, 49:629, 1987.

\bibitem{bag93}
C.~R. Bagshaw.
\newblock {\em Muscle Contraction}.
\newblock Chapman \& Hall, London, 1993.

\bibitem{ly_ta71}
R.~W. Lymn and E.~W. Taylor.
\newblock {\em Biochemistry}, 10:4617, 1971.

\bibitem{os_th95}
E.~M. Ostap and D.~D. Thomas.
\newblock {\em Biophys. J.}, 68:335s, 1995.

\bibitem{hux57}
A.~F. Huxley.
\newblock {\em Prog. Biophys.}, 7:255, 1957.

\bibitem{hil74}
T.~L. Hill.
\newblock {\em Prog. Biophys. Mol. Biol.}, 28:267, 1974.

\bibitem{ei_hi80}
E.~Eisenberg, T.~L. Hill, and Y.~Chen.
\newblock {\em Biophys. J.}, 29:195, 1980.

\bibitem{pa_co89}
E.~Pate and R.~Cooke.
\newblock {\em J. Muscle Res. Cell Motil.}, 10:181, 1989.

\bibitem{pa_wh93}
E.~Pate, H.~White, and R.~Cooke.
\newblock {\em Proc. Natl. Acad. Sci. USA}, 90:2451, 1993.

\bibitem{co_er92}
N.~J. C\'ordova, B.~Ermentrout, and G.~F. Oster.
\newblock {\em Proc. Natl. Acad. Sci. USA}, 89:339, 1992.

\bibitem{le_hu93}
S.~Leibler and D.~A. Huse.
\newblock {\em J. Cell. Biol.}, 121:1357, 1993.

\bibitem{de_vi97}
I.~Der\'enyi and T.~Vicsek.
\newblock {\em Biophys. J.}, 1997.
\newblock (submitted).

\bibitem{fi_si94}
J.~T. Finer, R.~M. Simmons, and J.~A. Spudich.
\newblock {\em Nature}, 368:113, 1994.

\bibitem{ra_ho93}
I.~Rayment, H.~M. Holden, M.~Whittaker, C.~B. Yohn, M.~Lorenz, K.~C. Holmes,
  and R.~A. Milligan.
\newblock {\em Science}, 261:58, 1993.

\bibitem{fi_sm95}
A.~J. Fisher, C.~A. Smith, J.~Thoden, R.~Smith, K.~Sutoh, H.~M. Holden, and
  I.~Rayment.
\newblock {\em Biophys. J.}, 68:19s, 1995.

\bibitem{co_bi79}
R.~Cooke and W.~Bialek.
\newblock {\em Biophys. J.}, 28:241, 1979.

\bibitem{pa_li92}
E.~Pate, M.~Lin, K.~Franks-Skiba, and R.~Cooke.
\newblock {\em Am. J. Physiol.}, 262:C1039, 1992.

\bibitem{co_pa85}
R.~Cooke and E.~Pate.
\newblock {\em Biophys. J.}, 48:789, 1985.

\bibitem{hil38}
A.~V. Hill.
\newblock {\em Proc. R. Soc. Lond.}, B126:136, 1938.

\bibitem{kr_sp86}
S.~J. Kron and J.~A. Spudich.
\newblock {\em Proc. Natl. Acad. Sci. USA}, 83:6272, 1986.

\bibitem{ha_no87}
Y.~Harada, A.~Noguchi, A.~Kishino, and T.~Yanagida.
\newblock {\em Nature}, 326:805, 1987.

\bibitem{sl_gl86}
J.~Sleep and H.~Glyn.
\newblock {\em Biochemistry}, 25:1149, 1986.

\bibitem{kat39}
B.~Katz.
\newblock {\em J. Physiol.}, 96:45, 1939.

\bibitem{hu_si71}
A.~F. Huxley and R.~M. Simmons.
\newblock {\em Nature}, 233:533, 1971.

\bibitem{ed_hw77}
K.~A.~P. Edman and J.~C. Hwang.
\newblock {\em J. Physiol.}, 269:255, 1977.

\bibitem{ajd94}
A.~Ajdari.
\newblock {\em J. Phys. I France}, 4:1577, 1994.

\bibitem{cs_fa97}
Z.~Csah\'ok, F.~Family, and T.~Vicsek.
\newblock {\em Phys. Rev. E}, 55:5179, 1997.

\bibitem{ju_pr95}
F.~J{\"u}licher and J.~Prost.
\newblock {\em Phys. Rev. Lett.}, 75:2618, 1995.

\bibitem{as_sc90}
A.~Ashkin, K.~Sch{\"u}tze, J.~M. Dziedzic, U.~Euteneuer, and M.~Schliwa.
\newblock {\em Nature}, 348:346, 1990.

\bibitem{de_vi95}
I.~Der\'enyi and T.~Vicsek.
\newblock {\em Phys. Rev. Lett.}, 75:374, 1995.

\bibitem{de_aj96}
I.~Der\'enyi and A.~Ajdari.
\newblock {\em Phys. Rev. E}, 54:R5, 1996.

\end{thebibliography}
\bibliographystyle{unsrt}

\end{document}